\documentclass[conference]{IEEEtran}

\ifCLASSINFOpdf
\else
\usepackage[dvips]{graphicx}
\fi
\usepackage[cmex10]{amsmath}

\begin{document}

\title{``Defective'' Logic:\\ Using spatiotemporal patterns in coupled
relaxation oscillator arrays for computation}

\author{\IEEEauthorblockN{Shakti N. Menon and Sitabhra Sinha}
\IEEEauthorblockA{The Institute of Mathematical Sciences,\\
CIT Campus, Taramani, Chennai 600113, India\\
Email: shakti@imsc.res.in, sitabhra@imsc.res.in}}

\maketitle

%=============================================================================%

\begin{abstract}
An intriguing interpretation of the
time-evolution of dynamical systems 
is to view it as a 
computation that transforms an initial state to a final one. This
paradigm has been explored in discrete systems such as cellular automata models,
where the relation between dynamics and computation has been
examined in detail. Here, motivated by microfluidic experiments on
arrays of chemical oscillators, we show that computation can be
achieved in continuous-state, continuous-time systems by using 
complex spatiotemporal patterns generated through a reaction-diffusion 
mechanism in coupled relaxation oscillators.
We present two paradigms that illustrate this computational
capability, namely,
using perturbations to (i) generate propagating configurations
in a system of initially exactly synchronized oscillators, 
and (ii) transform one time-invariant pattern to another. 
In particular, we have demonstrated a possible implementation
of NAND logic. This raises the possibility of universal computation
in such systems as all logic gates can be constructed from NAND gates.
Our work suggests that
more complex schemes can potentially implement arbitrarily
complicated computation using reaction-diffusion processes, bridging
pattern formation with universal computability.  
\end{abstract}

\IEEEpeerreviewmaketitle

%=============================================================================%

\section{Introduction}
%Physics and computation: It and Bit
The physical world abounds with dynamical systems in which the
temporal evolution of states results in processes as complex as
life. The usual approach for understanding the mechanisms underlying
such phenomena is to model them as a system of differential (or
difference) equations and to analyze the resulting spatiotemporal
patterns.
An alternative viewpoint is to consider these processes as
computations that transform input states into output states
iteratively. This suggests an intimate relation between the physical
world and that of information, an idea which has led to suggestions
that the universe itself may be viewed as a computer~\cite{Zuse1970}.
Indeed, the mutual connection between `it' and `bit'~\cite{Wheeler1989}
has been proposed
in light of the fact that
computational simulations may help provide insights
into the nature of physical laws~\cite{Feynman1982} while, in turn, 
physical principles provide constraints on the kind of computations that
are feasible~\cite{Zurek1990}.
These insights have inspired several studies on the connection between
spatiotemporal patterns generated by dynamical systems and
the computational processes that they might
represent~\cite{Crutchfield2012}, e.g., in models of neuronal
networks~\cite{Sinha1999}.
In this context, cellular automata is the most commonly studied class
of dynamical models~\cite{Crutchfield1995}, as their states, which form a
countably finite set, evolve in discrete space and time thus allowing
for an intuitive connection with existing abstract models of
computation, such as the Turing Machine~\cite{Lindgren1990}.

%-----------------------------------------------------------------------------%
\begin{figure}[!t]
\centering
\includegraphics[width=8.3cm]{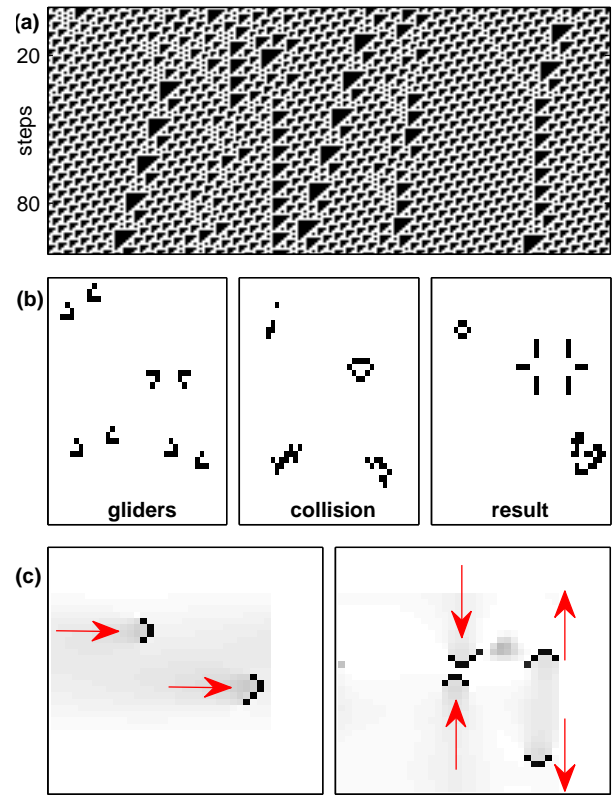}
\caption{(a) Spatiotemporal patterns in the dynamical evolution of
\emph{Rule~110}, a one-dimensional cellular automaton that has been shown to
be capable of universal computation~\cite{Cook2004}.
(b) Snapshots of the temporal evolution of \emph{Conway's ``Game of Life''},
a two-dimensional cellular automaton, showing (from left to right) different
stages of the collisions between propagating coherent structures known as
``gliders''. The use of gliders and more complex structures in the construction
of different types of logic gates has given weight to the claim that this
cellular automaton is capable of universal computation~\cite{Conway1982}.
(c) Propagating structures corresponding to phase defects in a two-dimensional
lattice of diffusively coupled relaxation oscillators~\cite{Singh2012}. The
left panel shows two coherent structures, resembling the gliders of (b), moving
horizontally in the direction indicated by the arrows, while the right panel
shows the interaction (via collision) between pairs of vertically moving
glider-like structures.
}
\label{fig1}
\end{figure}
%-----------------------------------------------------------------------------%

%Cellular automata as a model for studying the connection between It
%and Bit
While different types of cellular automata had been investigated and
classified based on the nature of the patterns manifested through their
spatiotemporal dynamics~\cite{Wolfram1983}, it has recently been shown
that they can also be classified in terms of their computational
complexity~\cite{Sinha2006}.
Certain classes of cellular automata have been shown to be capable of universal
computation~\cite{Cook2004}, a property 
which has been connected with the
existence of propagating coherent structures, such as
``gliders''~\cite{Conway1982} in the
two-dimensional cellular automata \emph{Conway's ``Game of Life''}
[Fig.~\ref{fig1}(a-b)]. 
%Reaction-diffusion systems in chemistry as an alternative paradigm
%for connecting pattern formation and computation (over reals)
Similar structures have recently been observed 
in spatially extended arrays of diffusively coupled relaxation oscillators that
model certain chemical systems~\cite{Singh2012}. These structures,
such as the propagating phase
discontinuities (`defects') in the array shown in  Fig.~\ref{fig1}~(c)
are visually analogous to the gliders in cellular automata.
This
brings up the intriguing possibility of using interactions between these
defects, and other spatiotemporal pattern regimes observed in this
system for computation. Indeed, the potential of computing through
chemical systems has been explored earlier, e.g., in the contexts of
enzyme-substrate reactions~\cite{Hjelmfelt1991}, waves in excitable
membranes~\cite{Steinbock1996} and bubbles in microfluidic
devices~\cite{Prakash2007}. However, the system we explore differs from
the earlier proposals in that oscillations (viz., in the
concentrations of the active species) play a fundamental role in
enabling computation. As interactions between biochemical oscillators
through reaction-diffusion mechanism~\cite{Goldbeter1997} 
is thought to be a critical
component for living systems~\cite{Goodwin1964}, the demonstration of computing via
chemical oscillations presents a direct connection between information
processing and the process of life.
%%%%%%%%%%%%%

In this article we introduce two paradigms for
implementing computation using local perturbations on the collective
dynamics of a model system of coupled relaxation oscillators. 
As the model system under consideration has close connections to the
experimental setup for microfluidic chemical oscillators, this
suggests
that the principles we
have presented
here can be used for constructing chemical logic gates~\cite{Epstein2007}.
%Here we report our investigation of a one-dimensional system.
Note that an important distinction between oscillator arrays
and cellular
automata lies in the fact that the former are usually 
described by systems of coupled
differential equations, and therefore have a continuous state space in
which
the system evolves in continuous time. Thus, computations implemented
in terms of chemical oscillations will be defined over a space of real
numbers~\cite{Blum1989} rather than a countably finite number of
states as in cellular automata. Furthermore, although glider-like
configurations can be shown in two-dimensional realizations of our
model system, here we focus on one-dimensional arrays for ease of
understanding. The demonstration of the capacity for computing even
in such simple systems underscores the potential inherent in
higher-dimensional realizations.

%=============================================================================%

\section{Model}
The spatiotemporal dynamics of an oscillatory chemical system is
investigated here by using a one-dimensional ring of $N$
relaxation oscillators, each diffusively coupled to its two nearest neighbors.
The local dynamics of individual oscillators are all identical, being described
by the generic FitzHugh-Nagumo (FHN) model which is a phenomenological
representation of the mechanism underlying periodic activity in many
chemical and biological
systems~\cite{Murray2002}. This model comprises a pair of dynamical variables
$u$ and $v$ that correspond to a fast activation and a slow inactivation
process, respectively, described by
\begin{equation}
\begin{split}
\dot{u} &= f(u, v) = u\ (1 -u)\ (u -\alpha) - v,\\
\dot{v} &= g(u, v) = \epsilon\ (k\ u -v -b),
\end{split}
\label{eq1}
\end{equation}
where $\alpha=0.139$, $k = 0.6$ are parameters describing the kinetics,
$\epsilon = 0.001$ characterizes the recovery rate and $b$ is a measure of the
asymmetry of the oscillator (measured as the ratio of the time spent by the
oscillator at high and low value branches of $u$). The parameter values are
chosen such that the system is in the oscillatory regime, and we have verified
that small variations in these values do not affect our results qualitatively.
The state of each oscillator at any given time is characterized by its
normalized phase angle
$\phi \in [0,1]$ that measures the position of the oscillator on its limit
cycle relative to the location of the peak of the activation variable
(considered to be the origin, i.e., $\phi=0$).

In chemical experiments involving oscillatory media, beads containing the
reactive solution are suspended in a chemically inert medium that allows
passage of only the inhibitory chemical species~\cite{Toiya2008}. Analogous to this, the
oscillators in our model are diffusively coupled via the inactivation variable
$v$. The dynamics of the resulting system is described by
\begin{equation}
  \begin{split}
  &\dot{u}_i        = f(u_i, v_i), \\
  &\dot{v}_i        = g(u_i, v_i) + D_v\ (v_{i-1}+v_{i+1}-2\ v_i),
  \end{split}
\label{eq2}
\end{equation}
where $i = 1, 2, \ldots, N$, and the ring connection topology is implemented
through the use of periodic boundary conditions ($v_0 = v_N$ and
$v_{N+1} = v_1$). The diffusion constant $D_v$ represents the strength of
coupling between neighboring relaxation oscillators that are connected
through their inactivation
variables. The dynamical equations are solved using an adaptive Runge-Kutta
scheme. To identify the dynamical regimes in the $b-D_v$ parameter space, the
system behavior is analyzed over many ($10^3$) realizations, with each
oscillator having a random initial phase, chosen from a uniform distribution.

%=============================================================================%

\section{Results}
\begin{figure}[!t]
\centering
\includegraphics[width=8.3cm]{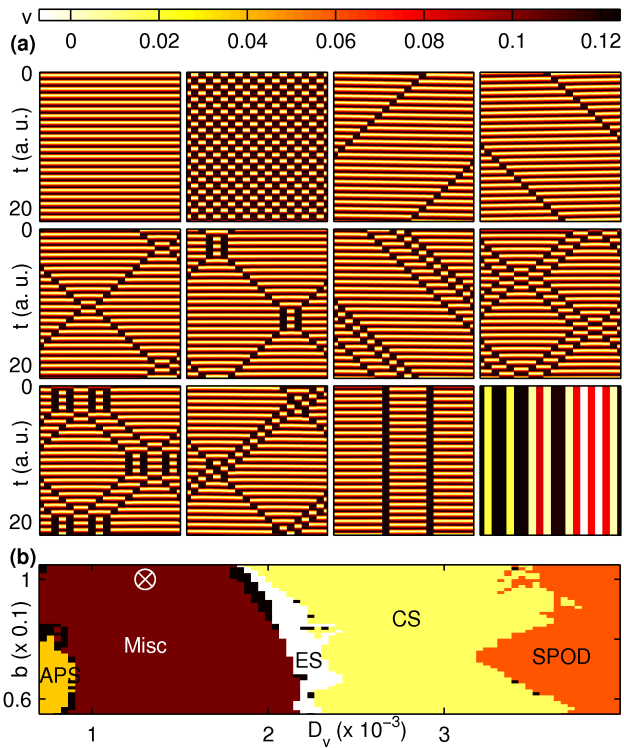}
\caption{
(a) Spatiotemporal evolution of a ring of $N=20$ relaxation oscillators, each
coupled diffusively to their nearest neighbors through the inactivation
variable. The individual oscillators are identical and described by the
FitzHugh-Nagumo equations (see text) with parameters $\epsilon=0.001$, $k=0.6$,
$b=0.1$ and $\alpha=0.139$. The strength of the diffusive coupling is
$D_v=1.13\times 10^{-3}$ for all panels except for the bottom right, in which
$D_v = 4\times 10^{-3}$ was used. The displayed patterns, obtained using
different initial states, are (left to right, top to bottom): Exact
Synchronization (ES), Anti-Phase Synchronization (APS), a Left-moving defect
(L), a Right-moving defect (R), a pair of Left and Right moving defects (LR)
that collide and pass through, a pair of Left and Right moving defects whose
collision results in a brief period of interaction before they pass through
(LR (long)), three defects moving in the same direction (3L/3R), two left and
two right moving defects (2LR) that collide and pass through, two left and two
right moving defects whose collisions result in a brief period of interaction
before they pass through (2LR (long)), moving defects other than those
described above (Others), two Static defects (2S) where a pair of
oscillators remain arrested in a stationary state, and a Spatially Patterned
Oscillator Death (SPOD) state. In each panel, time is expressed in arbitrary
units, normalized with respect to the period of the individual oscillators.
(b) Different dynamical regimes observed in a ring of $N=20$ oscillators in the
($D_v - b$) parameter plane, showing regions where the majority ($>50$\%) of
initial states result in ES, APS, SPOD, Chimera State (CS) and Miscellaneous
patterns (Misc). Regions colored black indicate parameter regimes where no
single class of patterns were observed for a majority of initial states. The
circled cross marks the location in parameter space where the dynamical
patterns (except SPOD) shown in the panels of (a) are obtained. The parameter
values for the individual oscillators are $\epsilon=0.001$, $k=0.6$ and
$\alpha=0.139$. 
}
\label{fig2}
\end{figure}

The system shows a range of spatiotemporal patterns including (a) Exact
Synchronization (ES) of all elements in the ring, i.e., all oscillators have
the same phase, (b) Anti-Phase Synchronization (APS) where oscillators next to
each other are in opposite phase, (c) Spatially Patterned Oscillator Death
(SPOD) where the oscillators are arrested in different stationary states, and
(d) Chimera States (CS), where oscillating elements coexist with elements
having nearly stationary behavior. In addition to these, several other possible
types of spatiotemporal patterns may be observed, and are categorized under a
Miscellaneous (Misc) class~\cite{Singh2012}. This includes several types of
propagating phase defects [Fig.~\ref{fig2}(a)]. Both APS and SPOD states have
been reported earlier in experiments on chemical systems~\cite{Toiya2008} and
it is likely that the other dynamical patterns seen in our simulations can also
be reproduced in appropriately designed experiments.

%-----------------------------------------------------------------------------%
%-----------------------------------------------------------------------------%

The robustness of these patterns are related to the size of their basins of
attraction (i.e., the fraction of randomly chosen initial conditions that lead
to a specific pattern) in the ($b-D_v$) parameter space [Fig.~\ref{fig2}(b)].
Distinct pattern regimes were identified using several order parameters
that are described in detail in Ref.~\cite{Singh2012}. Regions in parameter
space where ES, APS, SPOD, CS and Misc states are observed for the majority
of initial conditions (i.e., $>50$\% of the realizations) are labeled
correspondingly in Fig.~\ref{fig2}(b).
Over a small region of parameter space in the Misc regime of Fig.~\ref{fig2}(b),
we observe patterns characterized by one or more discontinuities of phase along
the oscillator ring (referred to as phase defects) that move against a
background of synchronized oscillations. The initial interactions between two
defects moving in opposite directions result either in their reflection or in
the annihilation of one or both of them. In the steady state, one finds that a
conserved number of defects can reflect off each other indefinitely.

%-----------------------------------------------------------------------------%
\begin{figure}[!t]
\centering
\includegraphics[width=8.3cm]{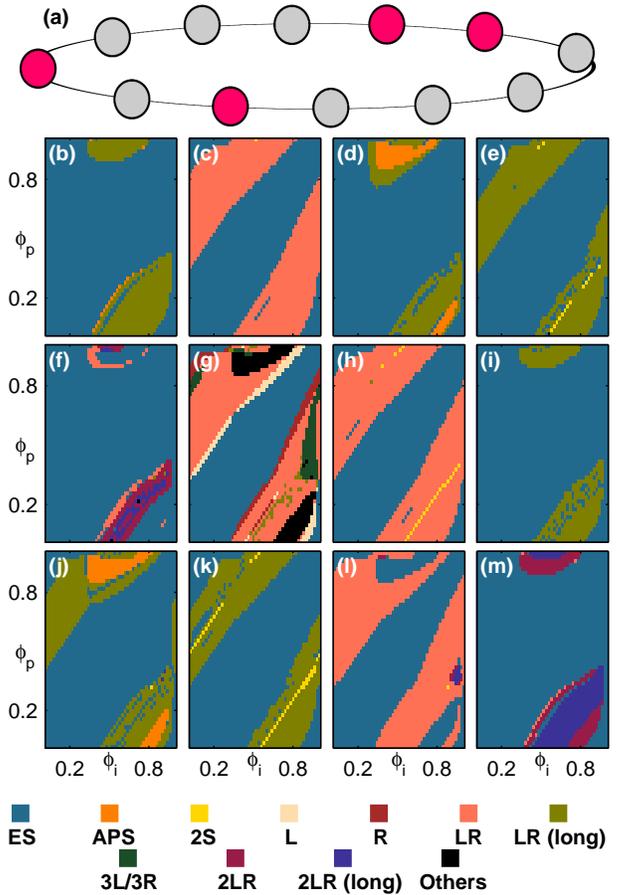}
\caption{(a) Schematic diagram illustrating the principle used to generate
different types of defects through local perturbations to an initial ES state. 
Each circle represents an oscillator that is coupled diffusively to its nearest
neighbor (via the inactivation variable). When the synchronized oscillators are
at some phase $\phi_i$ on the limit cycle, local perturbations are applied to
certain oscillators (shown in pink) by changing their phase to $\phi_p$, also
on the limit cycle. The perturbation scheme illustrated in this schematic can
have numerous representations because of the circular symmetry. One such
representation is $10100011\ldots$ (taking the bottom-most perturbed oscillator
as the origin and moving clockwise), where $0$ and $1$ correspond to
unperturbed and perturbed oscillators, respectively.
(b-m) Perturbation response ($\phi_i - \phi_p$) diagrams indicating the nature
of the defect patterns (represented using the color scheme shown below the
panels) obtained when a system of $N=20$ oscillators in an initial ES state is
perturbed according to the principle outlined in (a). Each oscillator is
identical, being described by the parameter values $\epsilon=0.001$, $k=0.6$,
$b=0.1$ and $\alpha=0.139$, and is coupled diffusively to each neighbour with
strength $D_v=1.13\times 10^{-3}$. The perturbation schemes used to generate
the response diagrams shown in the panels are: 
(b)~$1000\ldots$~, (c)~$1100\ldots$~, (d)~$1010\ldots$~, (e)~$1110\ldots$~,
(f)~$1001\ldots$~, (g)~$1011\ldots$~, (h)~$1111\ldots$~, (i)~$10001\ldots$~,
(j)~$10101\ldots$~, (k)~$11111\ldots$~, (l)~$101101\ldots$~, and
(m)~$10^{8}1\ldots$~, where $10^{k}1$ indicates that there are $k$ unperturbed
oscillators between two perturbed ones.
}
\label{fig3}
\end{figure}
%-----------------------------------------------------------------------------%

The resemblance of these propagating defects to the 
coherent structures in cellular
automata mentioned earlier suggests that the former may be used to implement a
form of computation. A possible approach is to perturb a system that is in an
ES state with a signal that acts as the input for a computation.
Specifically, this involves resetting the initial phase ($\phi_i$) of selected
oscillators in the ring to another phase ($\phi_p$). The sites that are
perturbed are indicated by $1$, while the unperturbed sites are indicated by
$0$. Thus, the perturbation pattern (or scheme) can be interpreted 
as a binary string
corresponding to the input signal [Fig.~\ref{fig3}~(a)].
The result of several different perturbation schemes, using
perturbation response ($\phi_i - \phi_p$) diagrams, are shown in 
Fig.~\ref{fig3}~(b-m). In these
diagrams, the nature of the defect patterns obtained when a selected set of
oscillators are perturbed from a specific initial phase $\phi_i$ to a given
phase $\phi_p$ are shown. 
The precise pattern is determined using heuristically derived order
parameters that utilize the qualitative differences between the
panels of Fig.~\ref{fig2}~(a), e.g., the total number of defects, the
number of ``long'' defects, the direction in which the defects travel,
etc.
Specific sequence structures in the perturbation
scheme are seen to yield predictable behavior in the response diagram. For
example, a scheme involving an alternation of $0$s and $1$s, such as
$1010\ldots$ and $10101\ldots$~, gives rise to large regions of APS in the phase
response diagram. On the other hand, a scheme that has a contiguous sequence
comprising an odd number of $1$s, such as $1110\ldots$ and
$11111\ldots$~, gives
rise to large regions of LR~(long).

%-----------------------------------------------------------------------------%
\begin{figure}[!t]
\centering
\includegraphics[width=8.3cm]{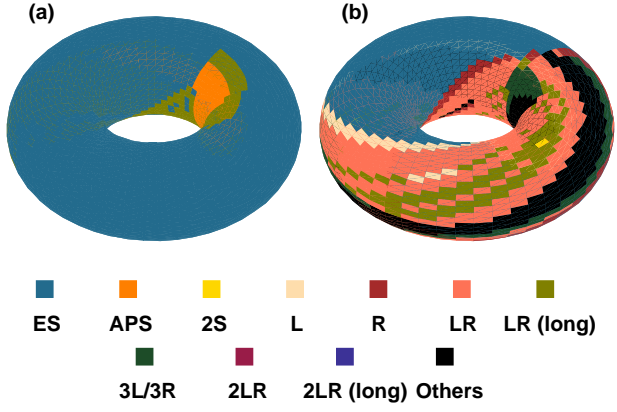}
\caption{Perturbation response ($\phi_i-\phi_p$) diagrams represented on a
two-dimensional torus for a ring of $N=20$ oscillators, initially in ES state,
diffusively coupled to their nearest neighbors (via the inactivation variable).
The nature of the ensuing defect patterns is indicated using the color scheme
shown below the panels. The toroidal representation of the response diagrams
arises naturally from the periodic nature of the phases $\phi_i$ and $\phi_p$
that correspond to the two axes of the toroid. The displayed figures, obtained
for perturbation schemes (a) $1010\ldots$ and (b) $101111\ldots$~, represent the
two topologically distinct classes of response diagrams that can be obtained
for all the perturbation schemes that have been investigated. These correspond
to the region in the ($\phi_i-\phi_p$) space where defects are
generated forming either (a) an isolated patch or (b) a continuous
ring that winds around the torus.
Each oscillator is identical, being
described by the parameter values $\epsilon=0.001$, $k=0.6$, $b=0.1$ and
$\alpha=0.139$, and is coupled diffusively to each neighbour with strength
$D_v=1.13\times 10^{-3}$.
}
\label{fig4}
\end{figure}
%-----------------------------------------------------------------------------%

While the response diagrams for different schemes differ greatly in their
detailed nature we observe that, for the schemes we have investigated, the
diagrams can be broadly classified into two topologically distinct
categories. This is shown in Fig.~\ref{fig4}, where the ($\phi_i -
\phi_p$) space is projected onto a torus, a representation that arises
naturally because of the periodicity of the phase angles $\phi_i$ and
$\phi_p$. In the first category, the region where
defects are generated by the perturbation
scheme is an isolated patch, while in the second, this region winds around the
torus to form a continuous ring. This can be alternatively interpreted as
follows: for one class of perturbation schemes, there exists a range of
given initial phases for which it is not possible to perturb the system away
from ES, while in the other class, for any initial phase $\phi_i$, it is
always possible to find a perturbed phase $\phi_p$ that will take the system
away from ES. From our results it appears that when the perturbation scheme
contains a subsequence `$11$' (i.e., any two neighboring elements are
perturbed) it belongs to the second class.

%-----------------------------------------------------------------------------%
\begin{figure}[!t]
\centering
\includegraphics[width=8.3cm]{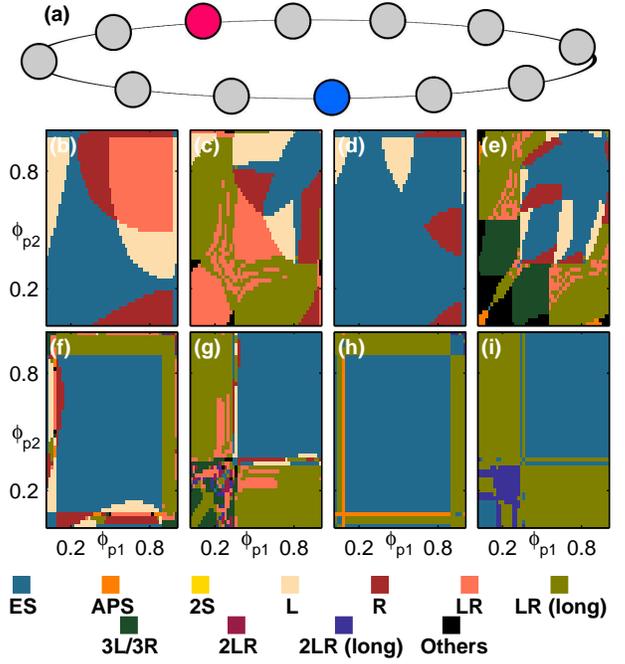}
\caption{(a) Schematic diagram illustrating the principle used to generate
different types of defects by perturbing any two sites in a ring of exactly
synchronized oscillators coupled diffusively to their nearest neighbors (via
the inactivation variable). When the synchronized oscillators are at some
phase $\phi_i$ on the limit cycle, different perturbations are applied to each
of the two oscillators (shown in pink and blue) by changing their phases to
$\phi_{p1}$ and $\phi_{p2}$, respectively. As in Fig.~\ref{fig3}~(a), the
perturbation scheme illustrated in this schematic can have multiple
representations because of the circular symmetry. One representation is
$10^{4}1\ldots$ (taking the bottom-most perturbed oscillator as the origin and
moving clockwise), where $0$ and $1$ correspond to unperturbed and perturbed
oscillators, respectively.
(b-i) Perturbation response ($\phi_{p1} - \phi_{p2}$) diagrams indicating the
nature of the defect patterns (represented using the color scheme shown below
the panels) obtained when a system of $N=20$ oscillators in an initial ES state
is perturbed according to the principle outlined in (a). Each oscillator is
identical, being described by the parameter values $\epsilon=0.001$, $k=0.6$,
$b=0.1$ and $\alpha=0.139$, and is coupled diffusively to each neighbour with
strength $D_v=1.13\times 10^{-3}$. The perturbation schemes used to generate the
response diagrams shown in each panel for a specific initial phase $\phi_i$ are:
(b) $11\ldots$ at $\phi_{i}=0.2$, (c) $11\ldots$ at $\phi_{i}=0.8$,
(d) $101\ldots$ at $\phi_{i}=0.2$, (e) $101\ldots$ at $\phi_{i}=0.8$,
(f) $10^{3}1\ldots$ at $\phi_{i}=0.4$, (g) $10^{3}1\ldots$ at $\phi_{i}=0.8$,
(h) $10^{9}1\ldots$ at $\phi_{i}=0.4$ and (i) $10^{9}1\ldots$ at
$\phi_{i}=0.8$. As in Fig.~\ref{fig3}, $10^{k}1$ indicates that there are
$k$ unperturbed oscillators between the two perturbed ones.
}
\label{fig5}
\end{figure}
%-----------------------------------------------------------------------------%

Using the above phenomenology, a possible principle for computation
can be realized by applying perturbations to
selected oscillators in the ring. In particular, we investigate the
implementation of a two-input logic gate by perturbing a
pair of elements that are initially in phase $\phi_{i}$ to phases
$\phi_{p1}$ and $\phi_{p2}$, respectively [Fig.~\ref{fig5}~(a)].
Interpreting low and high values of
$\phi_{p1}$ and $\phi_{p2}$ as $0$ and $1$, respectively, any combination of
two binary inputs can be implemented.
Fig.~\ref{fig5}~(b-i) shows the result of several different
perturbation schemes for given initial phases $\phi_i$, using perturbation
response ($\phi_{p1}-\phi_{p2}$) diagrams. 
From this, we note that the perturbation either results in the system
remaining in ES or transforming to a different state.
If we consider the output to
represent $1$ when the perturbation results in a final state
different from ES and $0$ if not, we observe that certain schemes, such as
$10^{3}1\ldots$ at $\phi_{i}=0.8$, correspond to NAND logic. 
This is because only a high-high combination of $\phi_{p1}$ and
$\phi_{p2}$ results in the system remaining in ES while the
other combinations (viz., low-low, low-high and
high-low) lead to defect states. As the NAND gate
is considered to be a universal one, i.e., it can be used in combination to
implement any other gate, this is a result of potential significance in using
relaxation oscillator array dynamics for computation.

%-----------------------------------------------------------------------------%
\begin{figure}[!t]
\centering
\includegraphics[width=8.3cm]{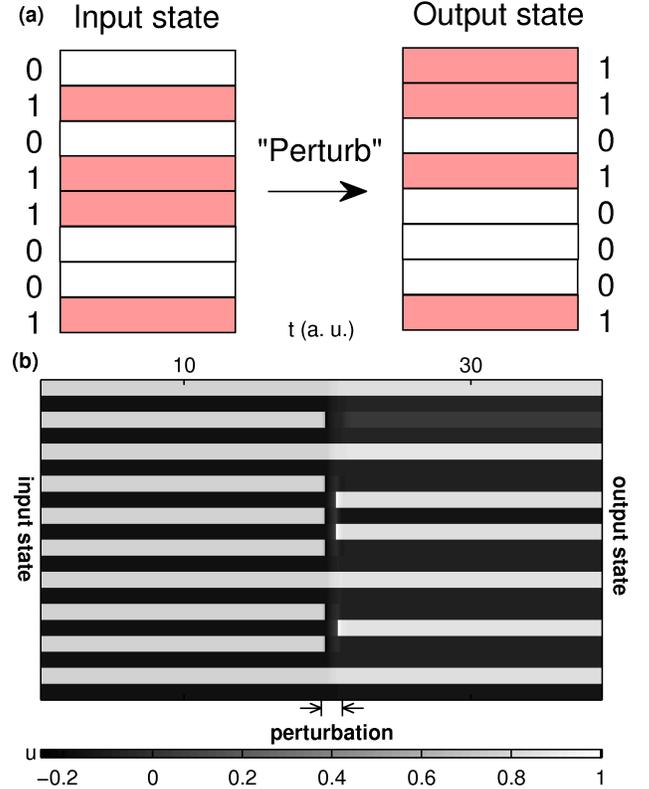}
\caption{(a) Schematic diagram illustrating a principle for computation
that involves the transformation of system states in the SPOD regime via local
perturbations, a specific example being shown in (b). The rows correspond to
individual oscillators, with the colors representing whether the corresponding
oscillator is arrested at a high (pink) or low (white) value of the activation
variable. This sequence of high and low values can be read as a string of $1$s
and $0$s, allowing us to map a particular state in the SPOD regime to a binary
sequence. In this representation, the initial state of the system corresponds
to an input binary string. Applying a local perturbation to specific 
oscillators results in a different
configuration of high and low values that can be interpreted as the output
binary string. Here, the activity following the perturbation represents
the computation that transforms the input to the output.
(b) The effect of local perturbations on the spatiotemporal evolution
in a ring
of $N=20$ oscillators. The perturbation involves stimulating
the inactivation component of six of the oscillators (corresponding to rows $3$,
$7$, $9$, $11$, $15$ and $17$, counting from top) for a short duration as
indicated in the figure. Subsequently, the system reorganizes into an altered
SPOD state. Thus, the perturbation can be interpreted as a computation that
transforms the input sequence $(10)^{10}$ to $10(001)^201(001)^30$. Note that
in the local spatial neighborhood of an oscillator, the perturbation is
functionally similar to a NOT logic gate. The individual oscillators in the
figure are identical, being described by the parameter values $\epsilon=0.001$,
$k=0.6$, $b=0.1$ and $\alpha=0.139$, and are coupled diffusively to their
nearest neighbors through the inactivation variable with strength
$D_v=5\times 10^{-3}$. Time is expressed in arbitrary units, normalized with
respect to the period of the uncoupled oscillators.
}
\label{fig6}
\end{figure}
%-----------------------------------------------------------------------------%

An alternative paradigm for implementing computation using the collective
dynamics of coupled relaxation oscillators becomes apparent when we focus on
the SPOD regime mentioned earlier. As oscillators are arrested in high or low
values of the activation variable in this regime, a natural interpretation of
the system state in terms of binary variables (viz., high=$1$, low=$0$)
suggests itself. Thus, a perturbation that transforms a specific sequence of
high and low values to another sequence can be interpreted as a computation
that generates an output binary string from a given input string
[Fig.~\ref{fig6}~(a)]. 
In order to demonstrate this principle, we use a specific perturbation
mechanism motivated by experiments involving light-sensitive
oscillating chemical systems, where oscillations can be terminated 
by increasing the intensity of incident
light~\cite{Vavilin1968}-\cite{Gaspar1983}. 
We implement this in our model [Eqs.~(\ref{eq1}) and (\ref{eq2})] by
introducing a parameter representing the intensity of the light
stimulus
inside the function $g(u,v)$, which governs the dynamics of 
the inactivation component $v$, similar to the formulation used in
Ref.~\cite{Krug1990}. A perturbation of the same given magnitude is
applied locally for a specific duration to a set of oscillators, which
results in a reconfiguration of the initial SPOD state to a different
one. An example is shown in Fig.~\ref{fig6}~(b), where the
perturbation can be interpreted as implementing a specific logic gate
or a
combination of gates in the neighborhood of the stimulated
oscillators.

%=============================================================================%

\section{Conclusion}
To conclude, we have shown that a simple model of coupled relaxation
oscillators that interact via diffusion of the inactivation component can give
rise to a variety of spatiotemporal patterns that can be potentially useful for
implementing computation. The generic nature of our model and its connection to
previous experiments suggest that the ideas outlined here can be implemented
using actual devices, for example, light-sensitive chemical
systems~\cite{Vavilin1968}-\cite{Gaspar1983}. Similarities between the
relaxation oscillator system that we have used and a chain of trapped ions,
as revealed by a recent theoretical study~\cite{Lee2011}, suggest 
another system that could be used as an experimental test-bed for our
results.
Our exploratory study of the use of perturbations to (i) generate propagating
configurations (involving phase defects) in a system of exactly synchronized
oscillators, and (ii) transform one time-invariant pattern to another, 
suggests
that these can be developed into schemes for implementing more
complicated computations. Our demonstration of NAND logic in one of
the implementations described above is of particular interest, as NAND
gates are considered to be universal gates due to the fact that all
possible logic gates can be constructed from combinations of NAND gates.
We note that our work connects pattern formation in reaction-diffusion
systems with the capacity for universal
computation and provides an intriguing link between the two enduring
legacies of Alan M. Turing~\cite{Turing1936}-\cite{Turing1952}.

%=============================================================================%

\section*{Acknowledgment}

The authors would like to thank Priyom Adhyapok for assistance in the initial
stages of the project, Rajeev Singh, S. Sridhar and V. Sasidevan 
for helpful discussions and
IMSc for providing access to the supercomputing cluster ``Satpura'', which is
partially funded by DST. This research was supported in part by the IMSc
Complex Systems Project.

%=============================================================================%
%=============================================================================%

%=============================================================================%
%=============================================================================%

% that's all folks
\end{document}